\documentclass[fleqn,usenatbib]{mnras}

\usepackage{mathptmx}

\usepackage[T1]{fontenc}

\newcommand\Tstrut{\rule{0pt}{2.6ex}}         
\newcommand\Bstrut{\rule[-0.9ex]{0pt}{0pt}}   

\usepackage{amssymb}	
\usepackage{graphicx}	
\usepackage{amsmath}	

\usepackage{hyperref}
\hypersetup{colorlinks=true,linkcolor=blue,citecolor=blue,filecolor=blue,urlcolor=blue}

\usepackage{siunitx}    

\usepackage[usenames, dvipsnames]{color}
\usepackage{soul}

\usepackage{subcaption}

\title[Evolution of radio-AGN feedback]{Cosmic evolution of radio-AGN feedback: confronting models with data}

\author[R. Kondapally et al.]{Rohit~Kondapally,$^{1}$\thanks{E-mail: rohitk@roe.ac.uk}
Philip~N.~Best,$^{1}$
Mojtaba~Raouf,$^{2}$
Nicole~L.~Thomas,$^{3,4}$
Romeel~Dav\'e,$^{1}$
\newauthor{Stanislav~S.~Shabala,$^{5}$
Huub~J.~A.~R\"{o}ttgering,$^{2}$
Martin~J.~Hardcastle,$^{6}$
Matteo~Bonato,$^{7}$}
\newauthor{Rachel~K.~Cochrane,$^{8}$
Katarzyna~Ma\l{}ek,$^{9,10}$
Leah.~K.~Morabito,$^{3,4}$
Isabella~Prandoni$^{7}$}
\newauthor{and Daniel.~J.~B.~Smith$^{6}$}
\\
$^{1}$SUPA, Institute for Astronomy, Royal Observatory, Blackford Hill, Edinburgh, EH9 3HJ, UK\\
$^{2}$Leiden Observatory, Leiden University, P.O. Box 9513, 2300 RA Leiden, The Netherlands\\
$^{3}$Centre for Extragalactic Astronomy, Department of Physics, Durham University, Durham DH1 3LE, UK\\
$^{4}$Institute for Computational Cosmology, Department of Physics, University of Durham, South Road, Durham DH1 3LE, UK\\
$^{5}$School of Natural Sciences, Private Bag 37, University of Tasmania, Hobart, TAS 7001, Australia\\
$^{6}$Centre for Astrophysics Research, Department of Physics, Astronomy and Mathematics, University of Hertfordshire, College Lane, Hatfield AL10 9AB UK\\
$^{7}$INAF-IRA, Via P. Gobetti 101, 40129, Bologna, Italy\\
$^{8}$Center for Computational Astrophysics, Flatiron Institute, 162 Fifth Avenue, New York, NY 10010, USA\\
$^{9}$National Centre for Nuclear Research, ul. Pasteura 7, 02-093 Warszawa, Poland\\
$^{10}$Aix Marseille Univ. CNRS, CNES, LAM, Marseille, France\\
}

\date{Accepted XXX. Received YYY; in original form ZZZ}

\pubyear{2022}

\begin{document}
\label{firstpage}
\pagerange{\pageref{firstpage}--\pageref{lastpage}}
\maketitle

\begin{abstract}
Radio-mode feedback is a key ingredient in galaxy formation and evolution models, required to reproduce the observed properties of massive galaxies in the local Universe. We study the cosmic evolution of radio-AGN feedback out to $z\sim2.5$ using a sample of 9485 radio-excess AGN. We combine the evolving radio luminosity functions with a radio luminosity scaling relationship to estimate AGN jet kinetic powers and derive the cosmic evolution of the kinetic luminosity density, $\Omega_{\rm{kin}}$ (i.e. the volume-averaged heating output). Compared to all radio-AGN, low-excitation radio galaxies (LERGs) dominate the feedback activity out to $z\sim2.5$, with both these populations showing a constant heating output of $\Omega_{\rm{kin}} \approx 4-5 \times 10^{32}\,\rm{W\,Mpc^{-3}}$ across $0.5 < z < 2.5$. We compare our observations to predictions from semi-analytical and hydrodynamical simulations, which broadly match the observed evolution in $\Omega_{\rm{kin}}$, although their absolute normalisation varies. Comparison to the Semi-Analytic Galaxy Evolution (\textsc{sage}) model suggests that radio-AGN may provide sufficient heating to offset radiative cooling losses, providing evidence for a self-regulated AGN feedback cycle. We integrate the kinetic luminosity density across cosmic time to obtain the kinetic energy density output from AGN jets throughout cosmic history to be $\sim 10^{50}\,\rm{J\,Mpc^{-3}}$. Compared to AGN winds, the kinetic energy density from AGN jets dominates the energy budget at $z \lesssim 2$; this suggests that AGN jets play an important role in AGN feedback across most of cosmic history. 
\end{abstract}




\section{Introduction}\label{sec:intro}
It is now widely believed that most galaxies in the Universe host a supermassive black hole (SMBH) at their centres. Significant evidence over the past two decades has shown the existence of a tight correlation between the mass of the SMBH and that of its galaxy's central disk \citep{2000ApJ...539L..13G,2001MNRAS.320L..30M}, and between the growth of the SMBHs and that of the galaxies \citep[e.g.][]{2013ARA&A..51..511K}. As these SMBHs accrete matter, during which they are known as active galactic nuclei (AGN), they release vast amounts of energy into their surroundings which can suppress star-formation and thereby regulate subsequent growth of their host galaxies \citep[e.g.][]{2006MNRAS.370..645B,2006MNRAS.365...11C}. This so-called AGN feedback effect is thought to result in the observed co-evolution between the SMBHs and their host galaxies \citep[e.g.][]{Best2005fagn,Best2006,2007ARA&A..45..117M,cattaneo2009_feedback_review,2012ARA&A..50..455F,2013ARA&A..51..511K,2014ARA&A..52..589H,Hardcastle2020}.

AGN feedback was initially introduced within semi-analytical models of galaxy formation and evolution to solve the cooling-flow problem and reproduce the observed high-mass end of the galaxy luminosity function \citep[e.g.][]{2006MNRAS.370..645B,2006MNRAS.365...11C}. This feedback from AGN was related to heating of the surrounding gas within the halo and preventing run-away star-formation, thereby limiting the growth of the most massive galaxies. This mode of feedback is commonly referred to as `radio-mode' feedback. The implementation of `radio-mode' feedback within semi-analytical models was motivated by observations of radio-jets from brightest cluster galaxies creating cavities in hot gas within massive haloes \citep[e.g.][]{Boehringer1993,Fabian2006}.

Feedback from AGN now forms a key ingredient in semi-analytical and hydrodynamical galaxy formation models, required to reproduce observed galaxy properties \citep[e.g.][]{2006MNRAS.370..645B,2006MNRAS.365...11C,2015MNRAS.446..521S,Croton2016,Kaviraj2017horizon_agn,Springel2018illustrius_tng,dave2019simba}. However, the implementation of the growth of black holes and AGN feedback models varies across different simulations. Moreover, the physical processes governing the growth of black holes and the formation and evolution of AGN jets operate on much smaller scales than the resolution of cosmological-scale hydrodynamical simulations, and are hence implemented using sub-grid prescriptions. The growth of black holes coupled with the growth and evolution of the gas and host galaxies can offer valuable insights, and when compared to observations can help constrain our understanding of black hole fuelling and feedback models.

Observationally, feedback from `radio-mode' AGN stems from two types of radio-loud AGN: high-excitation radio galaxies (HERGs) and low-excitation radio galaxies (LERGs), classified based on the nature of their optical emission line properties \citep[e.g.][]{2012MNRAS.421.1569B}. Detailed characterisation of these populations, primarily using wide-area radio surveys and optical spectroscopy, has revealed differences in both the properties of the central engine and the host galaxy properties. These observations show that LERGs have more massive black holes and tend to be hosted in more massive, quiescent, redder galaxies in richer environments than the HERGs \citep[e.g.][]{Best2006,Hardcastle2007,Tasse2008,Smolcic2009b,2012MNRAS.421.1569B,Janssen2012,Sabater2013,2014ARA&A..52..589H,Mingo2014,Tadhunter2016,Ching2017b,Croston2019,Hardcastle2020,Magliocchetti2022}. 

Studies of LERGs and HERGs in the nearby Universe have suggested that the differences in the observed properties of LERGs and HERGs may arise from differences in the Eddington-scaled accretion rates onto the SMBH \citep[e.g.][]{Best2005fagn,Hardcastle2007,Mingo2014,Delvecchio2017,Hardcastle2018,Hardcastle2020}. HERGs are thought to be undergoing radiatively-efficient accretion at high fractions of the Eddington-scaled accretion rates, typically from cold gas, leading to the formation of an optically-thick, geometrically thin accretion disk and torus structure \citep[e.g.][]{1973A&A....24..337S}. LERGs on the other hand are fuelled at low Eddington-scaled accretion rates by cooling hot gas within their haloes in an advection dominated accretion flow \citep[e.g.][]{1995ApJ...452..710N,Sharma2012,Gaspari2013}, and hence lack a radiatively-efficient accretion disk. However, recent deeper radio continuum surveys, probing over an order of magnitude fainter radio luminositites and higher redshifts, have found more overlap in key host galaxy and AGN properties between the two modes, suggesting that the picture may be more complicated \citep[e.g.][]{Whittam2018,Mingo2022,Kondapally2022,Whittam2022}.

The radio luminosity functions of the LERGs and HERGs evolve differently, with the LERGs dominating the space densities at lower luminosities compared to the HERGs at low redshifts \citep[e.g.][]{2012MNRAS.421.1569B,Gendre2013,Pracy2016}; this highlights the importance of studying both the LERG and HERG populations separately as the precise origins of the differences in fuelling and feedback between the HERGs and LERGs remain unclear. Determining the cosmic evolution of these AGN, their host galaxies, and their feedback effect are crucial in understanding their role in galaxy evolution across cosmic time. \citet{2014MNRAS.445..955B} were the first to study the evolution of LERGs and HERGs, separately, out to $z \sim 1$; they found that the HERGs showed a strong evolution with redshift, whereas the LERGs showed an overall mild evolution. Similar results were later found by \citet{Pracy2016} and \citet{Butler2019}. At low frequencies however, \citet{Williams2018} extended the analysis of the luminosity functions of LERGs across $0.5 < z \leq 2$ using 150\,MHz LOw Frequency ARray (LOFAR; \citealt{2013A&A...556A...2V}) observations of the Bo\"{o}tes field, finding a strong decline in their space densities with increasing redshift.

Recently, \citet{Kondapally2022} studied the evolution of the total radio-AGN and LERG luminosity functions out to $z \sim 2.5$ using the LOFAR Two-metre Sky Survey Deep Fields Data Release 1 (LoTSS-Deep DR1; \citealt{Tasse2021,Sabater2021,Kondapally2021,Duncan2021,Best2023}. LoTSS-Deep forms one of the deepest wide-field radio continuum surveys and covers a sky area of $\sim$ 25\,$\rm{deg^{2}}$ (where there is overlap with high-quality multi-wavelength data), and detects $> 11\,000$ radio-AGN (including $> 10\,000$ LERGs); this allowed them to probe much fainter radio luminosities than many previous studies \citep[e.g.][]{2014MNRAS.445..955B,Pracy2016,Williams2018,Butler2019} and study the low-luminosity AGN in detail while also tracing a wide range of galaxy environments and better constraining the bright-end of the luminosity function than deep narrow-area surveys \citep[e.g.][]{smolcic2017agn_evol_vla}. \citet{Kondapally2022} found that the LERG luminosity functions showed mild evolution across $0.5 < z \leq 2.5$ (the differences with the \citet{Williams2018} results were found to be largely due to differences in the source classification schemes employed; see \citealt{Kondapally2022}). When split by host galaxy type (quiescent versus star-forming), they found that the quiescent LERGs showed a strong decline in their space densities with increasing redshift; LERGs hosted by star-forming galaxies (SFGs) become more prominent across all luminosities at $z \gtrsim 1$. Moreover, they found that quiescent LERGs showed a roughly constant duty-cycle over the past $\sim$ 10\,Gyrs and that the strong negative evolution of the quiescent LERGs was in accordance with the space density of massive quiescent galaxies as their hosts. The radio-mode feedback models within simulations are often tuned to balance radiative cooling losses in massive quiescent hosts, in order to prevent gas cooling (and hence star formation) in these systems; therefore the different evolution seen from the different modes of radio-AGN has interesting implications for these feedback models.

In this paper, we study the cosmic evolution of radio-mode feedback by using the evolving radio luminosity functions to calculate the kinetic luminosity densities of radio-AGN and LERGs from LoTSS-Deep; this traces the volume integrated total heating output in the form of mechanical (kinetic) power of the radio jets. The observational measurements are then compared with predictions from hydrodynamical and semi-analytical galaxy formation models. The paper is structured as follows. In Section~\ref{sec:data} we describe the radio and multi-wavelength datasets used and the selection of radio-AGN and LERGs. Section~\ref{sec:heating_rate} presents a comparison of the cosmic kinetic luminosity density to other observations and simulations. In Section~\ref{sec:int_ke}, we calculate the total kinetic energy density output across cosmic history in the form of AGN jets to study the importance of jet AGN feedback. We present our conclusions in Section~\ref{sec:conclusions}. Throughout this work, we use a flat $\Lambda$CDM cosmology with $\Omega_{\rm{m}} = 0.3$, $\Omega_{\Lambda} = 0.7$, and $H_{\rm{0}} = 70~\rm{km\,s^{-1}\,Mpc^{-1}}$, and for calculating radio luminosities, a radio spectral index $\alpha = -0.7$ (where $S_{\nu} \propto \nu^{\alpha}$).

\section{Data}\label{sec:data}
\subsection{Radio and multi-wavelength data}\label{sec:obs_data}
The radio catalogues are taken from the LoTSS Deep Fields DR1 \citep{Tasse2021,Sabater2021}. The radio observations consist of deep repeated LOFAR High Band Antenna (HBA) observations (with an angular resolution of 6 arcsec) of the ELAIS-N1, Lockman Hole, and Bo\"{o}tes fields that reach an rms sensitivity of 20, 22, and 32\,$\rm{\mu Jy\,beam^{-1}}$ at 150\,MHz in the centres of each field, respectively. The LoTSS-Deep dataset forms the deepest wide-field radio-continuum survey at low-frequencies to date; this makes it ideal for studying the cosmic evolution of the faint radio-AGN population. The details of the radio calibration, imaging, and comparison to other radio-continuum surveys are presented by \citet{Tasse2021,Sabater2021}.

The three LoTSS-Deep fields were chosen due to the availability of deep, wide-field, multi-wavelength imaging \citep[see][and references therein]{Kondapally2021}, making these fields ideal for characterising the physical properties of the radio sources. In summary, the three LoTSS-Deep fields have coverage from the ultra-violet \citep[GALEX;][]{2007ApJS..173..682M}, optical (PanSTARRS Medium Deep Survey; \citealt{2016arXiv161205560C}, Hyper Suprime-Cam Subaru Strategic Program; \citealt{2019arXiv190512221A}, and the NOAO Deep Wide-Field Survey; \citealt{1999ASPC..191..111J}), near-infrared (UK Infrared Telescope Deep Sky Survey in ELAIS-N1 and Lockman Hole; \citealt{2007MNRAS.379.1599L}, and \textit{J}, \textit{H}, \textit{K$_{\rm{s}}$} data in Bo\"{o}tes from \citealt{gonzalez2010newfirmbootes}), mid-infrared (Spitzer IRAC and MIPS surveys; \citealt{2003PASP..115..897L,2012PASP..124..714M,eisenhardt2004iracshallow,2009ApJ...701..428A}, and far-infrared (Herschel; \citealt{oliver2012hermes}) wavelengths.

The identification of the host galaxies of the radio-detected sources, and the association of radio components (where the radio source finder had not correctly grouped physical sources in the catalogue) was performed by \citet{Kondapally2021}. The host galaxies were identified using a combination of the statistical likelihood ratio method \citep{1977A&AS...28..211D,1992MNRAS.259..413S} and a visual classification scheme \citep[see also][]{2019A&A...622A...2W}, whereas the source association was performed using visual classification only. While the radio data covers a much larger area ($\sim 68\,{\rm{deg^{2}}}$ in each field), the host galaxy identification process was limited to the areas in each field with the best available multi-wavelength data; this process resulted in host galaxies for $>$ 97 per cent of the radio-detected sources, with the final radio catalogue consisting of 81\,951 radio sources, across $\sim 25\,{\rm{deg^{2}}}$ over the three fields \citep{Kondapally2021}. Photometric redshifts for the radio sources (and the underlying multi-wavelength catalogues) were generated by \citet{Duncan2021} using a hybrid machine learning and template-based approach optimised for deep radio-continuum surveys \citep[see][]{duncan2018photz_templates,duncan2018photz_ml}.

\subsection{SED fitting and source classification}\label{sec:sedfit}
The multi-wavelength photometry and photometric redshifts were used to perform spectral energy distribution (SED) fitting for all of the radio-detected sources\citep{Best2023}. For each source, the SED fitting process was performed using four different routines: \textsc{AGNFitter} \citep{CalistroRivera2016}, Bayesian Analysis of Galaxies for Physical Inference and Parameter Estimation \citep[\textsc{bagpipes};][]{Carnall2018}, Code Investigating Galaxy Evolution \citep[\textsc{cigale};][]{Burgarella2005,Noll2009,Boquien2019}, and Multi-wavelength Analysis of Galaxy Physical Properties \citep[\textsc{magphys};][]{daCunha2008}. 

The output of this SED fitting process was used to perform source classifications and derive stellar masses and star-formation rates (SFRs) for the full radio dataset. In summary, \citet{Best2023} used fits to the AGN accretion disk and torus models present in both \textsc{AGNFitter} and \textsc{cigale} to define a parameter, `AGN fraction', which represents the fraction of the mid-infrared emission arising from AGN as compared to the host galaxy components. This `AGN fraction', along with a comparison of the goodness of fit from the SED fitting codes with and without AGN components, were used to identify the so-called `optical' (SED) AGN (also known as radiative-mode AGN). For such sources, the \textsc{cigale} stellar masses and SFRs were adopted, as these were found to have lower scatter than those from \textsc{AGNfitter}, while the masses estimated from \textsc{magphys} or \textsc{bagpipes} for these sources may be inaccurate due to the lack of an AGN component in these codes. For sources that did not show signs of a radiative-mode AGN, the average of the \textsc{bagpipes} and \textsc{magphys} results were generally used to derive `consensus' stellar masses and SFRs as these codes include better sampling of the range of potential stellar populations. Then, to identify the radio-AGN, \citet{Best2023} selected sources whose radio emission predominantly arose from the AGN by identifying sources that showed an excess of $> 0.7\,\rm{dex}$ ($\sim 3\sigma$) in radio luminosity, over the level expected from star-formation processes alone based on the consensus SFRs and SFR-radio luminosity relation for star-forming galaxies \citep[e.g.][]{gurkan2018lofar_sfr,Smith2021}.

HERGs are the radio-loud subset of the radiative-mode AGN population (i.e. also display signs of an AGN in their SED); these are identified as sources showing radio-excess and an SED AGN. The LERGs are identified by the presence of radio-jets only; these were hence identified as sources that display a radio-excess AGN but were not selected as AGN based on their SEDs. The radio luminosity functions for these LERG and HERG samples were constructed and discussed by \citet{Kondapally2022}.

In this study, we follow \citet{Kondapally2022} in selecting quiescent galaxies using the specific star-formation rates (sSFRs) of galaxies, such that sources which satisfy $\rm{sSFR} < 0.2/t_{\rm{H}}(z)$ are defined as quiescent galaxies, where $t_{\rm{H}(z)}$ is the age of the Universe at redshift $z$; this criterion was found to be broadly consistent with quiescent galaxies selected using rest-frame UV colour-colour diagrams \citep[e.g.][]{Pacifici2016,Carnall2018,Carnall2020}. In total, across the redshifts $0.5 < z \leq 2.5$ analysed in this study, our sample consists of 9485 radio-excess AGN, of which 8409 are LERGs (with 2974 of these LERGs being hosted in quiescent galaxies; hereafter quiescent LERGs or Q-LERGs).

We note that the identification of radiative-mode AGN by \citet{Best2023} may be incomplete (leading to HERGs being mis-classified as LERGs), especially in the absence of X-ray data (which is not available in ELAIS-N1 and Lockman Hole). To test this, we utilised the X-Bo\"{o}tes Chandra survey \citep{Kenter2005}; this dataset was also used to identify X-ray detected AGN during the source-classification process in the Bo\"{o}tes field \citep[see][]{Duncan2021,Best2023}. We find that only $\sim 5\%$ of the radiative-mode AGN in Bo\"{o}tes were identified as AGN using X-ray data alone. Therefore, while the lack of wide-field X-ray observations in ELAIS-N1 and Lockman Hole may result in a small fraction of mis-classifications, \citet{Best2023} found that the fraction of sources classified as LERGs and HERGs across the three fields were consistent with each other. Furthermore, \citet{Kondapally2022} also constructed the LERG luminosity functions in each of the three LoTSS-Deep fields separately, finding good agreement across the fields. These results indicate that the lack of X-ray data in ELAIS-N1 and Lockman Hole does not have a significant effect on the derived luminosity functions and results in this paper.

\section{Evolution of radio-AGN feedback over cosmic time}\label{sec:heating_rate}
The observed 150\,MHz radio luminosity functions (LFs) for the radio-AGN in LoTSS-Deep were presented by \citet{Kondapally2022}. The LFs were calculated using the $1/V_{\rm{max}}$ method and the evolution of the radio-excess AGN, LERGs, Q-LERGS (i.e. LERGs hosted in quiescent galaxies) was characterised. The observed LFs and the best-fitting evolution models derived by \citet{Kondapally2022} are shown in Appendix~\ref{sec:lf_comp}, with the radio-excess AGN LFs recomputed over different redshift bins for the analysis in this paper. In this section, we consider the implications of the evolution of the observed radio-AGN LFs from \citet{Kondapally2022} on the evolution of the amount of energy deposited into their host galaxies by the radio-AGN across cosmic time. We also perform a comparison with simulations in physical space by using scaling relations to convert observed radio luminosities into jet kinetic powers.

\subsection{Kinetic powers of radio-AGN}\label{sec:kinetic_lum}
The kinetic energy carried in the jets can provide a significant energetic output in the form of work done on the surrounding environment; this so-called mechanical (kinetic) power from the jets can be orders of magnitude larger than the monochromatic radio luminosities. The jet kinetic power is often estimated from observed radio luminosities by using a scaling relation. One such relation often used in literature \citep[e.g.][]{smolcic2017agn_evol_vla,Butler2019} was derived by \citet{Willott1999} who determined the kinetic powers by using minimum energy arguments to estimate the energy stored within lobes, and combined this with estimates of radio source lifetimes and energy losses from inflating the radio source. Uncertainties in our knowledge of the physics of radio jets and their composition, along with departures from the minimum energy condition result in significant uncertainties in the calibration of this relation; \citet{Willott1999} combined all these uncertainties into a single parameter, $f_{\rm{W}}$, with typical values in the range of $f_{\rm{W}} \sim 1 - 20$. Converted to 1.4\,GHz luminosity \citep{2014ARA&A..52..589H}, this relation is given as:
\begin{equation}\label{eq:p_sync}
    L_{\rm{kin,\,sync}} = 4 \times 10^{35} \left(f_{\rm{W}}\right)^{3/2} \left( L_{\rm{1.4\,GHz}}/10^{25}\,\rm{W\,Hz^{-1}} \right)^{0.86}\,\rm{W},
\end{equation}
where $L_{\rm{1.4\,GHz}}$ is the 1.4\,GHz radio luminosity, and $f_{\rm{W}}$ is the uncertainty parameter on the calibration of the scaling relation.

Another method of calculating jet kinetic powers is based on studying the cavities in the hot gas created as the radio jets plough through the surrounding material, which can be observed in the X-rays \citep{Boehringer1993}. The jet kinetic power can be estimated from the radio luminosity by considering the energy required to expand the lobes and inflate the X-ray cavities of pressure $p$ with a volume $V$, $E_{\rm{cav}} = f_{\rm{cav}}\, pV$, where $f_{\rm{cav}} = 4$ is the commonly adopted value \citep[e.g.][]{Cavagnolo2010,2014MNRAS.445..955B,Pracy2016,Butler2019}, corresponding to the $pV$ of work done, and $3pV$ of energy stored in the relativistic particles; a good correlation is seen between the cavity powers and 1.4\,GHz radio luminosities \citep[e.g.][]{Birzan2004,Cavagnolo2010,Timmerman2022}. \citet{2014ARA&A..52..589H} derived a best-fit relation, largely based on the results from \citet{Birzan2008,Cavagnolo2010}, given as
\begin{equation}\label{eq:l_kin_fcav}
L_{\rm{kin,\,cav}} = 7 \times 10^{36} f_{\rm{cav}} \left( L_{\rm{1.4\,GHz}} / 10^{25}\, \rm{W\,Hz^{-1}} \right)^{0.68} \rm{W}.
\end{equation}
The normalisation of this relation is found to be in good agreement with that of the \citet{Willott1999} relation when using $f_{\rm{W}} = 15$ and $f_{\rm{cav}} = 4$ \citep[e.g.][]{2014ARA&A..52..589H,smolcic2017agn_evol_vla}, however the \citet{2014ARA&A..52..589H} relation has a shallower slope which will result in higher jet powers at low luminosities.

It is important to note that the above relations have a large scatter that is dominated by the systematic effects and assumptions about the unknown physics of the radio sources. Therefore a simple conversion between radio luminosity and kinetic power is likely not appropriate; even for jets of a consistent kinetic power, the radio luminosity varies over the lifetime of a radio source, and depends on the energy density and magnetic field of the radio lobes and hence the environment into which the radio lobes are expanding, and is also influenced by the assumed spectral index \citep[see][]{Shabala2012,Shabala2013,Hardcastle2013,Godfrey2016,Croston2018,Hardcastle2018_sim,Hardcastle2019}. However, while a relation based on radio luminosity alone may not be accurate for inferring jet kinetic powers for individual sources, we are interested in the heating output from the AGN at a population level; the use of either of the above scaling relationships for typical values of the uncertainty parameters (see below) should therefore provide a reasonable mean value.

\begin{figure*}
    \centering
    \includegraphics[width=\textwidth]{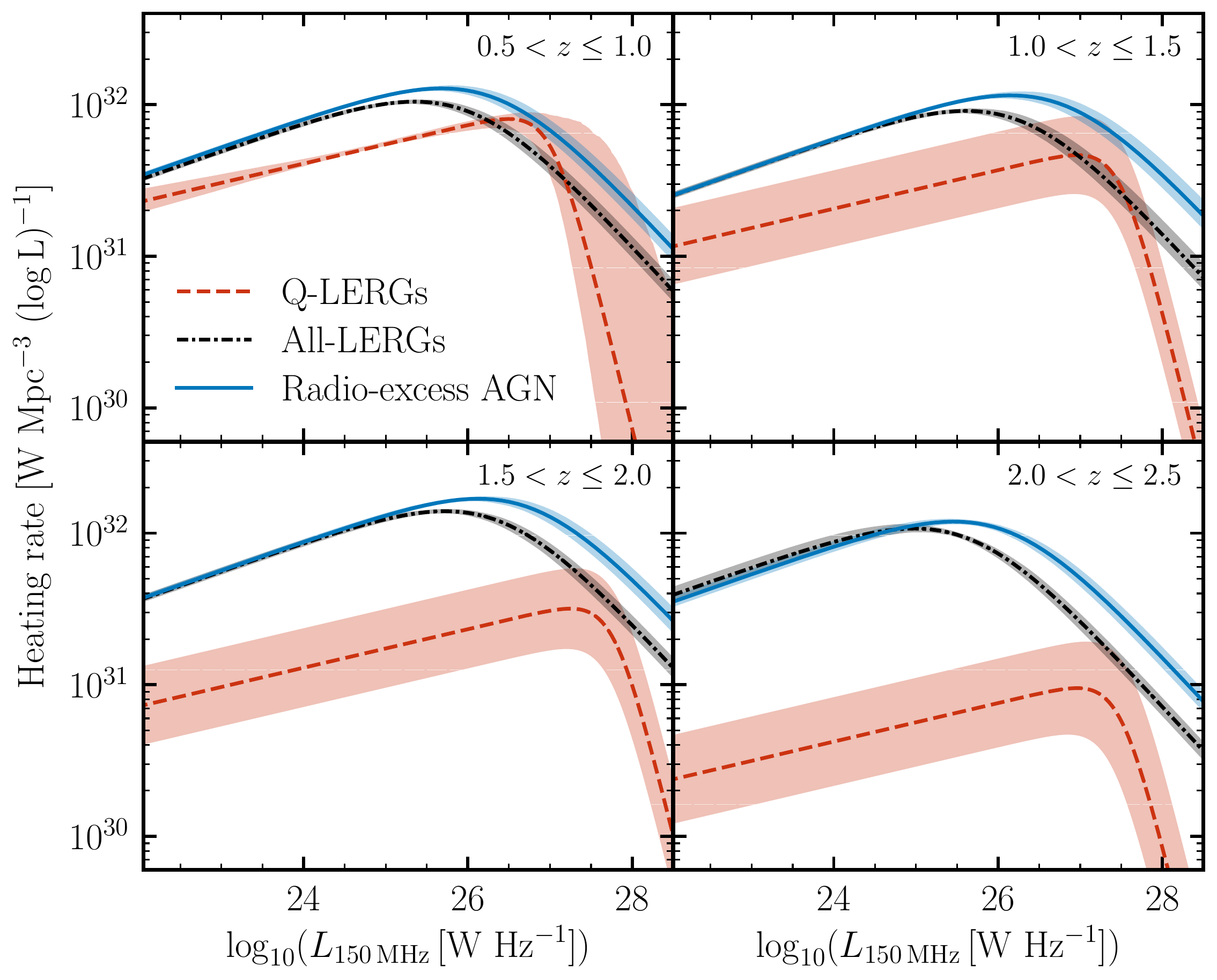}
    \caption{\label{fig:hrate_l150}Cosmic evolution of the heating rate as a function of radio luminosity in four redshift bins for LERGs (black), Q-LERGs (red), and radio-excess AGN (blue). These are calculated by convolving the best-fit LFs with the jet kinetic powers (estimated using equation~\ref{eq:l_kin_fcav}), with the shaded regions showing uncertainties on the best-fitting LF models. The heating rate function peaks at the break of the luminosity function, and therefore most of the heating output comes from relatively high luminosity sources.}
\end{figure*}

An additional issue is that the above scaling relation has been determined from characterisation of sources at low-redshifts and it is possible that the radio luminosity of the sources of a given jet power (and hence the above relation) will evolve with redshift (e.g. due to cosmic evolution of magnetic field strengths, or increasing inverse-Compton losses); a well-constrained relation out to high redshifts is, however, still lacking. Recently, \citet{Hardcastle2019} used the projected linear source sizes, redshifts, and radio luminosities to infer the jet kinetic powers of resolved AGN out to $z \sim 0.7$ from the first data release of the wide-area LoTSS survey \citep{2019A&A...622A...1S} using the dynamical model from \citet{Hardcastle2018_sim}. High-resolution imaging using the LOFAR international baselines for LoTSS-Deep \citep[e.g.][]{Sweijen2022} will provide more robust sizes and help extend the jet power inference models to higher redshifts (Hardcastle et al. in prep.). 

In the analysis that follows, we use the \citet{2014ARA&A..52..589H} relation (Equation~\ref{eq:l_kin_fcav}) with $f_{\rm{cav}} = 4$ and assume that this local relation is applicable over the full redshift range studied, but note that this might lead to systematic errors. In Appendix~\ref{ap:lkin_rels}, we study the impact of using different scaling relations for estimating jet powers on our results. We find that the volume-integrated heating rate across cosmic time predicted from these two relations (Equations~\ref{eq:p_sync}~and~\ref{eq:l_kin_fcav}) is in good agreement (see Fig.~\ref{fig_ap:hrate_z_var}), while the use of some other relations in the literature would also produce almost identical trends in the cosmic evolution of the volume-integrated heating rate, but with typical changes to the overall normalisation by around 0.3\,dex. This normalisation uncertainty needs to be borne in mind when comparing the output of the simulations with the observational data.

\begin{figure*}
    \centering
    \includegraphics[width=\textwidth]{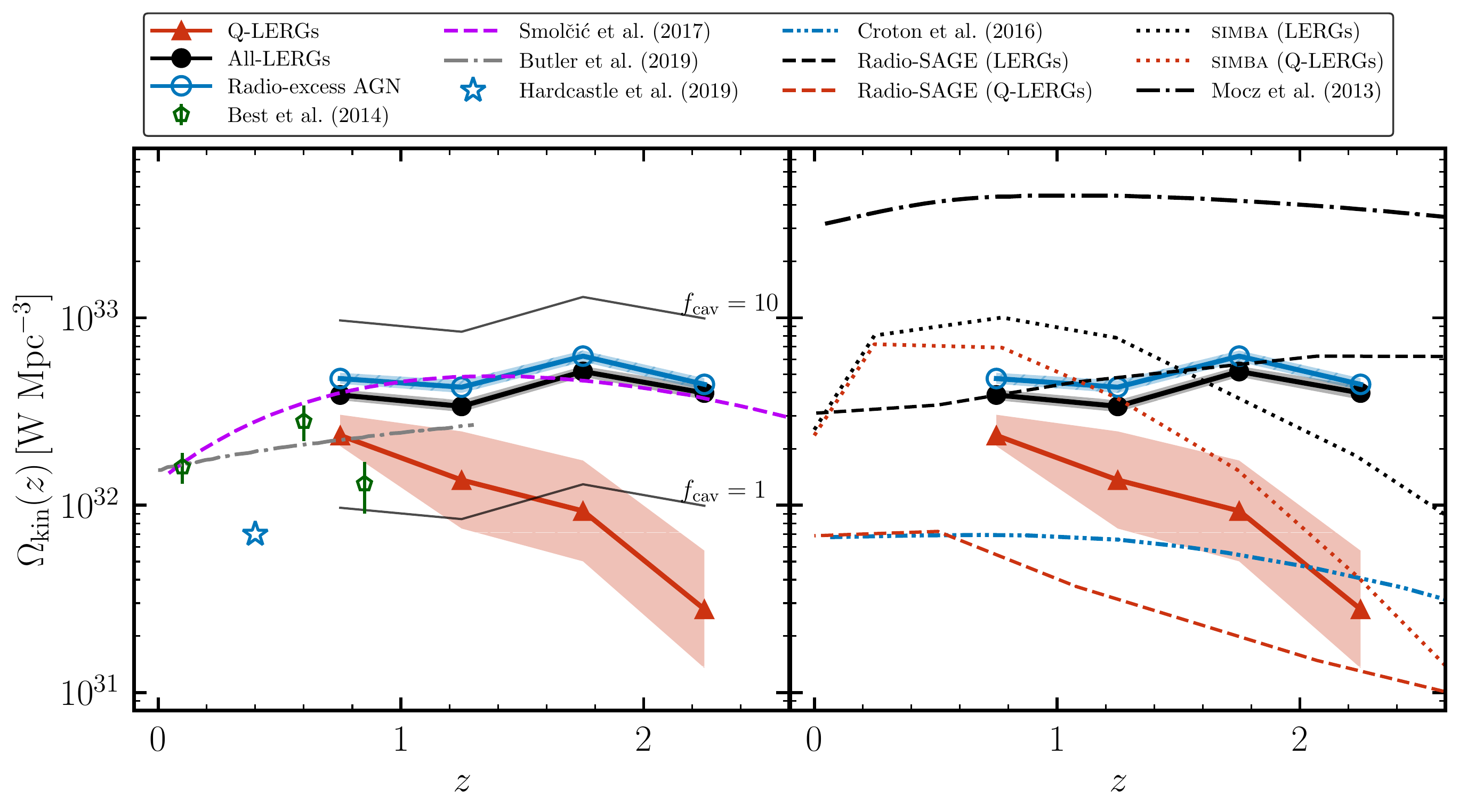}
    \caption{\label{fig:lerg_hrate}The evolution of the kinetic luminosity density, $\Omega_{\rm{kin}}(z)$ compared to observations in literature (\textit{left}) and to simulations (\textit{right}). Measurements for all LERGs (black points), Q-LERGs (red triangles), and radio-excess AGN (blue open circles) in the LoTSS Deep Fields are shown on both panels. The shaded regions in each case represent the uncertainties on the kinetic luminosity density based on uncertainties in fitting the evolution of the LFs. The kinetic luminosity density was calculated using equation~\ref{eq:l_kin_fcav} with $f_{\rm{cav}} = 4$ and the best fit LDE models for the three AGN populations (see Appendix~\ref{sec:lf_comp}). The black lines show the systematic shift that would be obtained for the total LERG heating rate using the values $f_{\rm{cav}} = 1$ and $f_{\rm{cav}} = 10$ that account for the systematic uncertainties in the estimate of kinetic powers. Observational results for the LERG population from \citet{2014MNRAS.445..955B} and \citet{Butler2019}, along with radio-excess AGN from \citet{smolcic2017agn_evol_vla}, and radio-AGN from \citet{Hardcastle2019} are shown in the left panel. Our measurements are compared to predictions from \textsc{sage} \citep{Croton2016}, \textsc{radio-sage} \citep{Raouf2017}, \textsc{simba} \citep{dave2019simba,Thomas2021}, and \citeauthor{Mocz2013} (\citeyear{Mocz2013}; their LERG-equivalent population) in the right panel.}
\end{figure*}

\subsection{Cosmic evolution of the kinetic luminosity density}\label{sec:hrate_evol}
The evolving radio luminosity functions of AGN from \citeauthor{Kondapally2022} (\citeyear{Kondapally2022}; also shown in Appendix~\ref{sec:lf_comp}) can be combined with equation~\ref{eq:l_kin_fcav} to estimate the kinetic heating rate as a function of radio luminosity (also known as the specific heating rate function), given as
\begin{equation}\label{eq:hrate_logL}
\Psi(L_{\rm{1.4\,GHz}}, z) = L_{\rm{kin}}(L_{\rm{1.4\,GHz}}) \times \rho\left(L_{\rm{1.4\,GHz}}, z\right) \, \rm{W\,Mpc^{-3}\,logL^{-1}}.
\end{equation}
Using the above best-fitting models for the evolution of the LFs, we calculated the specific kinetic heating rate for the radio-excess AGN, the total LERG, and the Q-LERG populations separately. For each population, the best-fitting LF models were shifted to 1.4\,GHz using the standard spectral index $\alpha = -0.7$ to obtain $\rho(L_{\rm{1.4\,GHz}}, z)$ for each redshift bin; this was then convolved with $L_{\rm{kin}}(L_{\rm{1.4\,GHz}})$ (calculated  using equation~\ref{eq:l_kin_fcav}) to determine the kinetic heating rate as a function of radio luminosity. The resulting curves for the radio-excess AGN, LERGs, and Q-LERGs are shown in Fig.~\ref{fig:hrate_l150} with the shaded regions corresponding to uncertainties in the modelling of the evolution of the LFs. For all three populations, the heating rate function peaks at high radio luminosities, near the break in the LFs; the location of this peak occurs at slightly higher luminosities at higher redshifts. At $z < 1$, the majority of the heating rate output from LERGs comes from the Q-LERGs across all radio luminosities, as would be expected given the LFs observed in Fig.~\ref{fig:lerg_lf_comp}. The heating rate from Q-LERGs peaks at higher radio luminosities than that of the other populations. This may in part be due to the uncertainty in modelling the break luminosity in the LFs for the LERGs and radio-excess AGN; for Q-LERGs, the characteristic space density at each redshift is fixed based on the available host galaxies, resulting in one less free parameter at each redshift interval. At higher redshifts, while the heating rates from LERGs show little evolution in shape and normalisation, the heating output from Q-LERGs decline sharply, in line with the observed evolution of the LFs.

The specific heating rate function from Eq.~\ref{eq:hrate_logL} can be integrated to estimate the total kinetic luminosity density (also known as the kinetic heating rate), $\Omega_{\rm{kin}}(z)$. This is given, in units of $\rm{W\,Mpc^{-3}}$, as
\begin{equation}\label{eq:kin_density_int}
\Omega_{\rm{kin}}(z) = \int \Psi(L_{\rm{1.4\,GHz}}, z) \, \rm{d}\log L_{\rm{1.4\,GHz}}\, \rm{W\,Mpc^{-3}}.
\end{equation}
For each redshift bin in Fig~\ref{fig:hrate_l150}, the specific heating rate function is integrated as a function of the radio luminosity down to $10^{-4}\,L_{\star}(z)$ to obtain the redshift evolution of the integrated kinetic luminosity density for each of the total LERGs (black circles), the Q-LERGs (red triangles), and the radio-excess AGN (blue open circles), as shown in Fig.~\ref{fig:lerg_hrate}. In each case, the shaded region corresponds to the uncertainties in the kinetic luminosity density based on uncertainties in fitting the evolution of the LFs. The narrow black lines show the systematic shift that would be produced for the total LERGs based on the extreme values of $f_{\rm{cav}} = 1$ and $f_{\rm{cav}} = 10$; this illustrates the effect of uncertainties in the calibration of the radio-luminosity to jet kinetic power relationship on the absolute values obtained; changes due to different calibration relations discussed in Appendix~\ref{ap:lkin_rels} are typically $\lesssim 0.3\,\rm{dex}$, so smaller than these extreme values of $f_{\rm{cav}}$.

The kinetic luminosity density for both the radio-excess AGN and the LERGs remains roughly constant between $0.5 < z \leq 2.5$ with $\Omega_{\rm{kin}} \approx 4-5 \times 10^{32}\,\rm{W\,Mpc^{-3}}$ for both populations at $z = 0.75$; this indicates that the dominant source of heating from the radio-loud population comes from the LERGs (rather than the HERGs) across redshift, and that at least out to $z \sim 2.5$, the importance of feedback from LERGs is broadly uniform. In contrast, for the Q-LERG population, the kinetic luminosity density decreases steadily with increasing redshift, from $\Omega_{\rm{kin}} \approx 2.7 \times 10^{32}\,\rm{W\,Mpc^{-3}}$ at $z = 0.75$ to nearly an order of magnitude lower by $z \sim 2.5$.

In Fig.~\ref{fig:lerg_hrate} (left), we compare our observations with other measurements of the kinetic luminosity densities from the literature. The estimates obtained for the `jet-mode AGN' population by \citet{2014MNRAS.445..955B} for $z < 1$, also derived using the \citet{2014ARA&A..52..589H} jet power scaling relation with $f_{\rm{cav}} = 4$ are shown by green symbols.  The results from \citet{2014MNRAS.445..955B} show an increase out to $z \sim 0.6$ and then show a decrease in the kinetic luminosity density, resulting in better agreement with our Q-LERG population, within errors, than the total LERG population. One potential reason for the better agreement with the quiescent-LERGs could be due to the method by which the `jet-mode AGN' were selected by \citet{2014MNRAS.445..955B}; these were identified using emission line ratio diagnostics to select sources with relatively low emission line fluxes or equivalent widths from [\ion{O}{II}] or [\ion{O}{III}] lines. Jet-mode AGN hosts with considerable star-forming activity will result in higher equivalent widths of these spectral lines and therefore the spectroscopic classification used by \citet{2014MNRAS.445..955B} may result in a sample similar to the quiescent-LERG population identified in this study.

The grey dash-dotted line illustrates the kinetic luminosity density for the LERGs obtained using a pure density evolution model for the evolution of the LFs as determined by \citet{Butler2019} out to $z \sim 1.3$, using the \citet{Cavagnolo2010} relation. The results from \citet{Butler2019} show a steady increase with increasing redshift and agree well with our estimates for the total LERG population, even when extrapolating their results to higher redshifts. We note that even if the \citet{2014MNRAS.445..955B} selection is similar to our Q-LERGs, the good agreement between the \citet{2014MNRAS.445..955B} and \citet{Butler2019} results is expected at $z < 0.6$, where the total LERG population is dominated by Q-LERGs. The dashed purple line shows the results for all radio-AGN from \citet{smolcic2017agn_evol_vla} determined using the \citet{Willott1999} relation with $f_{\rm{W}} = 15$; their results agree well with our measurements for the evolution of the radio-excess AGN kinetic luminosity density across redshift.

We also compare to results from \citet{Hardcastle2019}, who used data from LoTSS DR1 to calculate the jet kinetic powers for radio-AGN by incorporating the projected source sizes using the analytic model from \citet{Hardcastle2018_sim}. \citet{Hardcastle2019} used their inferred jet powers to calculate the jet kinetic luminosity function at $z < 0.7$ and integrated this to find $\Omega_{\rm{kin}} \sim 7 \times 10^{31}\,\rm{W\,Mpc^{-3}}$ (shown as a blue star in Fig.~\ref{fig:lerg_hrate}). This result is systematically lower (by a factor of $\sim 2- 3$) than other observations shown in Fig.~\ref{fig:lerg_hrate} at this redshift, however this offset is found to be due to the different methods of estimating jet kinetic powers. Similarly, \citet{Turner2015} found a lower kinetic luminosity density using their dynamical model than the results presented in this work. Unlike \citet{Hardcastle2019}, our study and those by \citet{2014MNRAS.445..955B,smolcic2017agn_evol_vla,Butler2019} use radio luminosity to jet-power scaling relationships which generally predict higher jet powers at low luminosities \citep[see fig.~A4 of][and Appendix~\ref{ap:lkin_rels}]{Hardcastle2019}. We note that if we use the \citet{Willott1999} scaling relationship with $f_{\rm{W}} = 4$ instead, which will result in lower jet powers at low luminosities compared to the \citet{2014ARA&A..52..589H} relation adopted in this study, our measurement for the kinetic luminosity density of radio-excess AGN would match well with the results from \citet{Hardcastle2019}.

\subsection{Comparison of kinetic luminosity density with simulations}\label{sec:hrate_sims_comp}
In Fig.~\ref{fig:lerg_hrate} (right), we compare the observed kinetic luminosity densities with predictions from recent simulations. We first compare with predictions from the Semi-Analytic Galaxy Evolution (\textsc{sage}) model \citep{Croton2016}, which provides an enhancement to the previous model by \citet{2006MNRAS.365...11C}. \citet{2006MNRAS.365...11C} considered the prescription of SMBH growth and AGN feedback in two classes: ``quasar mode'' and ``radio mode'' in their terminology. We used the predictions of the black hole accretion rate density ($\dot{m}_{\rm{BH,R}}$) over cosmic time for the ``radio-mode'' population from \citet[][see their fig.~3]{2006MNRAS.365...11C}. The $\dot{m}_{\rm{BH,R}}(z)$ were then translated into a luminosity of the black hole assuming $L_{\rm{BH}} = \eta \dot{m}_{\rm{BH,R}} c^{2}$, where $\eta = 0.1$ is the standard black-hole accretion efficiency and $c$ is the speed of light. This luminosity generated from the accretion process is considered equivalent to the kinetic luminosity density output from the LERGs derived in this study. As part of an update to this model, \citet{Croton2016} introduced a ``radio mode efficiency'' parameter, $\kappa_{\rm{R}}$ within \textsc{sage} to provide a more realistic  treatment of AGN feedback cycle by attempting to couple the heating provided by the AGN with the cooling. As a result, in addition to $\eta = 0.1$, \citet{Croton2016} scale their $\dot{m}_{\rm{BH,R}}(z)$ and hence the black hole luminosities by $\kappa_{\rm{R}} = 0.08$ (which we also apply here).

The resulting prediction from the \textsc{sage} model is shown in the right panel of Fig.~\ref{fig:lerg_hrate} (dot-dot dashed blue line). The radio-excess AGN measurements show a similar shape to the \textsc{sage} model prediction, but offset to higher values by nearly an order of magnitude. This could be due to calibration errors in the scaling relation, or due to some fraction of the heating from radio-AGN (in particular from very extended sources) being deposited on scales larger than that useful to offset cooling losses. Moreover, radio-mode feedback in the \textsc{sage} model is scaled to a level required to provide AGN heating that can offset radiative cooling, however, at higher redshifts in particular, the LERG activity and heating output occurs predominantly in star-forming LERGs; the feedback cycle from LERGs within star-forming galaxies remains unclear. The kinetic luminosity density of the Q-LERG population is also systematically higher than the \textsc{sage} model at all but the highest redshift bin; these results therefore suggest that the energy output by the Q-LERG population alone is sufficient to offset the radiative cooling losses within \textsc{sage} at least out to $z \sim 2.5$, thereby providing evidence for a self-regulating AGN feedback cycle in quiescent galaxies hosting a LERG.

We then compare our observations to predictions from another semi-analytical model, \textsc{radio-sage} \citep{Raouf2017}, that builds upon the \textsc{sage} model to focus on intermittent AGN jet activity, providing an improved modelling of radio-mode feedback \citep{Raouf2017}. The accretion rate of matter onto the black hole is approximated by the Bondi-Hoyle relation \citep{Bondi1952}, with two accretion states implemented in \textsc{radio-sage}. The `hot-mode' accretion occurs at low Eddington-scaled accretion rates ($f_{\rm Edd} \equiv \frac{\dot{M}_{\rm BH}}{\dot{M}_{\rm BH,Edd}} < \alpha_{\rm{crit}}$), resulting in an advection dominated accretion flow \citep{1994ApJ...428L..13N,1995ApJ...452..710N}, and `cold-mode' (or radiatively-efficient) accretion occurs at high rates ($f_{\rm{Edd}} > \alpha_{\rm{crit}}$) resulting in the formation of an optically-thin accretion disk \citep{1973A&A....24..337S}. The `hot-mode' (i.e. with low Eddington-scaled accretion rates)  sources are taken to be analogous to the LERGs following observed differences in the accretion rate properties of LERGs and HERGs \citep[e.g.][]{2012MNRAS.421.1569B}; here, we used $\alpha_{\rm{crit}} = 0.03$ to select LERGs from \textsc{radio-sage} \citep{Shabala2009}. The quiescent galaxies in \textsc{radio-sage} are selected using the same sSFR criterion as the observations, with the SFR calculated based on the average over the past $\sim 280$\,Myr (the time-step duration between each redshift snapshot). The output jet power as a result of this accretion is then estimated following $L_{\rm{kin,hot}} = \eta \dot{M}_{\rm{BH}} c^{2}$, where c is the speed of light and $\eta$ is the accretion efficiency that is constrained in \textsc{radio-sage} to be 0.35 using observations as described by \citet{Raouf2017}. 

The \textsc{radio-sage} predictions for the LERG and Q-LERG populations (dashed black and red lines, respectively) are also shown in Fig.~\ref{fig:lerg_hrate}. The integrated kinetic luminosity densities for LERGs from \textsc{radio-sage} are in excellent agreement with our observations, showing a gradual increase with redshift. The predictions for Q-LERGs decline with redshift, showing a similar slope to the observations but with a normalisation that is lower by up to $\sim 0.5\,\rm{dex}$; this suggests that the star-forming galaxies in \textsc{radio-sage} play an important role in radio-AGN feedback at all times. Given the good agreement with the total LERG population, this also implies that \textsc{radio-sage} may not match the observed space density of the Q-LERGs, with a significantly higher fraction of the heating output from LERGs being performed in star-forming galaxies within \textsc{radio-sage}; we will investigate this directly by comparing the radio LFs and other host galaxy properties from simulations in future work. In addition, \textsc{radio-sage} is able to model large sources which heat gas out to well beyond the cooling radius; therefore more jet power is required for the same amount of feedback within \textsc{radio-sage} compared to the more idealised feedback efficiency model in \textsc{sage}. The observed size distribution of sources, using high-resolution LOFAR data, can be compared to predictions from \textsc{radio-sage} in a future paper.

We also compare our results to predictions from the \textsc{simba} cosmological hydrodynamical simulation, which incorporates a unique sub-grid prescription for black hole growth and feedback \citep{dave2019simba}. \textsc{simba} employs a two-mode accretion model, with the accretion from hot gas ($T > 10^{5}\,\rm{K}$; i.e. hot-mode) described by Bondi accretion \citep{Bondi1952}, and the accretion from cold gas ($T < 10^{5}\,\rm{K}$; i.e. cold-mode) being described by a torque-limited accretion model \citep{Hopkins2011,Angles-Alcazar2017}. Following \citet{Thomas2021}, who studied the properties of radio-AGN within \textsc{simba}, LERGs are identified as sources where the Bondi accretion (or accretion via hot gas) dominates the black hole accretion rate (based on the dichotomy in accretion rates observed for the LERGs and HERGs; \citealt{2012MNRAS.421.1569B}). This classification is performed based on the average accretion rate (from both hot and cold gas) over the past 50\,Myr period. Subsequently, quiescent galaxies (and hence Q-LERGs) in \textsc{simba} are identified using the same sSFR criterion as in the observations (see Sect.~\ref{sec:obs_data}). Following the prescription of \citet{Kording2008} used by \citet{Thomas2021} to estimate radio-source observables, the jet kinetic powers for the LERGs and Q-LERGs within \textsc{simba} are calculated as $L_{\rm{kin}} = 0.2\,L_{\rm{bol}}$, where $L_{\rm{bol}} = 0.1 \dot{M}_{\rm{BH}}c^{2}$ is the bolometric luminosity of the AGN, and $\dot{M}_{\rm{BH}}$ is the black hole accretion rate density.

The predictions from \textsc{simba} for both the LERGs and Q-LERGs (dotted black and red lines, respectively) are also shown in Fig.~\ref{fig:lerg_hrate}. For LERGs, \textsc{simba} predicts $\Omega_{\rm{kin}}$ that increases from $z=0$ and peaks at $\Omega_{\rm{kin}} \approx 10^{33}\,\rm{W\,Mpc^{-3}}$ by $z \sim 1$ before declining at higher redshifts. Although these predictions show a different redshift evolution than our observations, displaying higher $\Omega_{\rm{kin}}$ values at lower redshifts and lower values at higher redshifts, the \textsc{simba} predictions for LERGs are typically within a factor of two of our observations. The Q-LERGs within \textsc{simba} show a similar slope to the observations, matching them at high redshifts but producing more heating output by a factor of $\sim 3$ at $z \lesssim 1.5$ than observed. These results suggest that in order to reproduce the massive galaxy population at $z=0$, \textsc{simba} requires more heating from AGN jets than observed at $z \lesssim 1.5$ (as determined for the observations using the \citet{2014ARA&A..52..589H} jet power scaling relationship).

Finally, we also show the prediction for the evolution of radio-mode feedback from \citet{Mocz2013} who develop a model for the evolution and energetic output of supermassive black holes based on accretion only. Using the derived Eddington luminosity distributions from this model, they predict the kinetic luminosity density for different modes of AGN, with the LERG equivalent population being referred to as the `low kinetic' (LK) mode AGN. Their LK mode AGN results are displayed in Fig.~\ref{fig:lerg_hrate} (black dash-dotted line), which shows a similar evolution to the observed LERGs but with a systematically higher normalisation by up to an order of magnitude. The \citet{Mocz2013} results are also higher than other observations and similar studies that directly use radio LFs and a scaling relationship for kinetic powers to compute the kinetic luminosities \citep[e.g.][]{Kording2008,Merloni2008}. \citet{Mocz2013} propose that this systematic offset is due to faint sources missing from the observed radio LFs and due to a fraction of bright, older sources that may fall below the flux limit due to the decrease in radio luminosity as a source ages. The latter effect, however, is also incorporated in the \textsc{radio-sage} simulation, where the dynamics and radio luminosity evolution of sources across their lifetimes are modelled, which shows a much lower normalisation that is in agreement with our observations. Furthermore, as indicated in Fig.~\ref{fig:hrate_l150}, the contribution of low luminosity sources to the integrated kinetic luminosity density is not significant.

\begin{figure}
    \centering
    \includegraphics[width=\columnwidth]{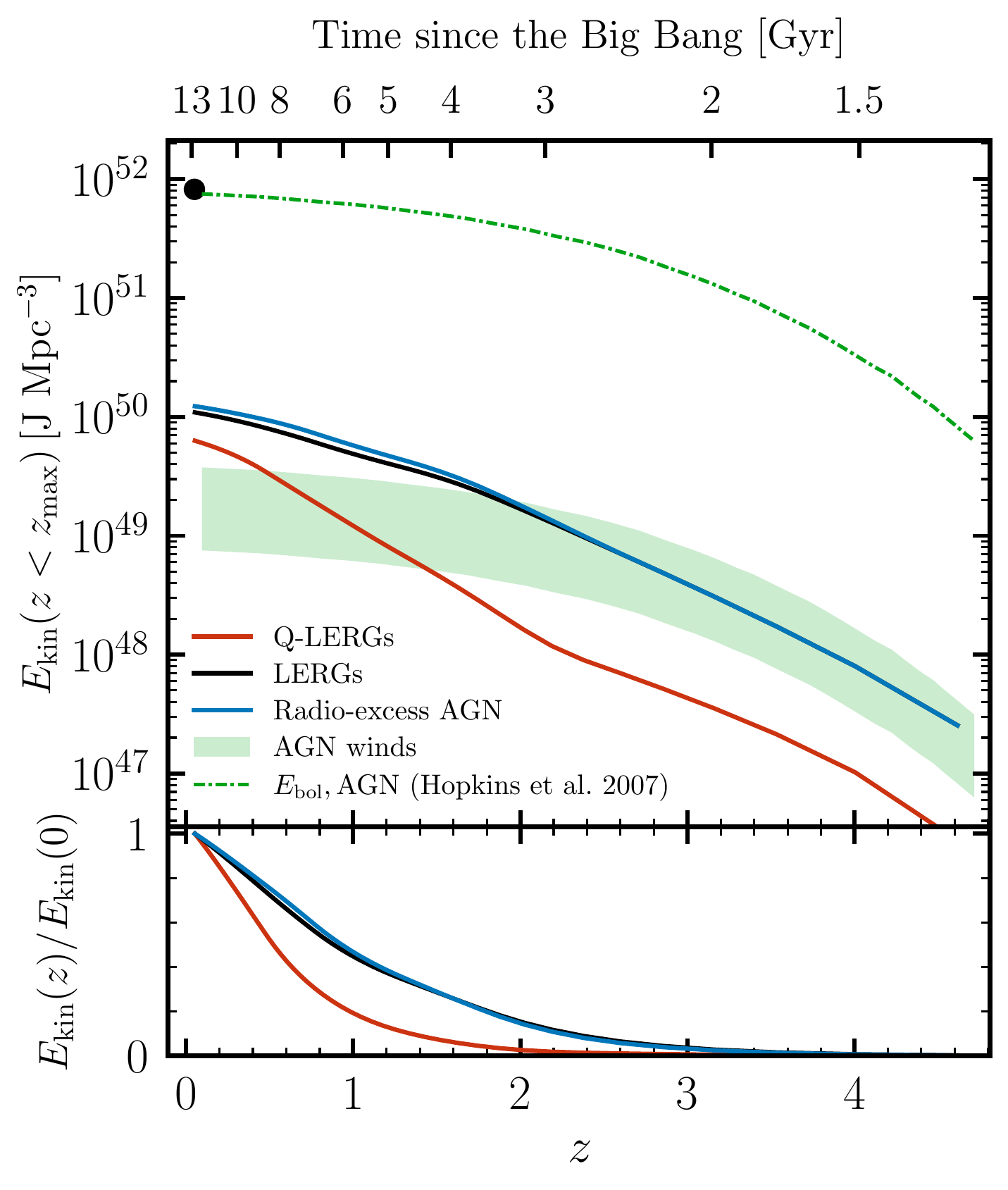}
    \caption{\label{fig:ke_dens_tot}\textit{Top:} The cumulative kinetic energy density radiated from AGN in different forms across cosmic time (in units of $\rm{J\,Mpc^{-3}}$). The observed kinetic luminosity densities for Fig.~\ref{fig:lerg_hrate} are integrated from $z=5$ to $z=0$, using extrapolation beyond the coverage of the dataset (see Sect.~\ref{sec:int_ke}) to determine the kinetic energy output from AGN jets for the three groups of radio-AGN. The circle shows the total energy radiated by BHs using the local BH mass density from \citet{Marconi2004} and an accretion efficiency of 10\%. The green line shows the bolometric energy radiated from quasars calculated using the black hole mass densities (and accretion efficiency of 10\%) that was determined by \citet{Hopkins2007}. The green shaded region shows the range of possible values for kinetic energies from AGN driven winds determined by scaling the \citet{Hopkins2007} line (see text). Our results indicate that while AGN winds may be more important at early times, AGN jets play a more important role in AGN feedback at $z \lesssim 2$. \textit{Bottom:} The fraction of the total (i.e. $z=0$) kinetic energy density emitted across cosmic time.}
\end{figure}

\section{Kinetic energy density from AGN jets across cosmic history}\label{sec:int_ke}
We can study the total kinetic energy output from the radio-AGN in the form of jets by integrating the kinetic luminosity density across cosmic time from Fig.~\ref{fig:lerg_hrate}. To do this, we extrapolated the observed $\Omega_{\rm{kin}}$ to higher redshifts (i.e. $z > 2.5$) by assuming that our results follow the evolution observed by \citet{smolcic2017agn_evol_vla} out to $z_{\rm{max}} \sim 5$ (the approximate redshift limit of measurements made by \citealt{smolcic2017agn_evol_vla}) for the radio-excess AGN and LERGs; for the Q-LERGs, due to the strong evolution with redshift, we linearly extrapolated our measurements of $\Omega_{\rm{kin}}(z)$ for this population out to $z_{\rm{max}}=5$. We extrapolated our results to lower redshifts ($z < 0.5$) for all three groups of AGN by again using the values from \citet{smolcic2017agn_evol_vla}; at these low redshifts, the LERGs will dominate the total radio-excess AGN population (as shown by the agreement between the \citet{smolcic2017agn_evol_vla} and the \citet{2014MNRAS.445..955B} LERG data at $z < 0.7$), with the Q-LERGs dominating the total LERG population. The kinetic energy density output from the AGN jets was then obtained by integrating these extrapolated kinetic luminosity density curves with respect to time.

In Fig.~\ref{fig:ke_dens_tot}, we show the cumulative integral (with respect to time) of these extrapolated kinetic luminosity density curves from $z=5$ to $z=0$ in units of $\rm{J\,Mpc^{-3}}$. The results for radio-excess AGN, LERGs, and Q-LERGs are shown by the same colours as in Fig.~\ref{fig:lerg_hrate}. The total time-integrated kinetic energy density in the form of jets from radio-AGN (and for the LERG subset given their similarities in the kinetic heating rates) across cosmic history is $E_{\rm{kin}} \approx 10^{50}\,\rm{J\,Mpc^{-3}}$, with $\sim$ 50 per cent of this emitted within the past $\sim 6\,\rm{Gyr}$; this agrees well with other measurements \citep[e.g.][]{2023Galax..11...21H}. For Q-LERGs, the corresponding value is $E_{\rm{kin}} \sim 7 \times 10^{49}\,\rm{J\,Mpc^{-3}}$, but they deposit most of their total kinetic energy at late times, $z \lesssim 0.5$ (Fig.~\ref{fig:ke_dens_tot}; \textit{bottom}).

For comparison, the green line shows the bolometric radiative energy density from AGN estimated by using an accretion efficiency of 10\% and the evolution of the black hole mass density from \citet{Hopkins2007} who determined this by integrating their quasar luminosity function using the Soltan argument \citep{Soltan1982}; the present day value of this agrees well with integrated measurements of the local black hole mass density from \citet[][black circle]{Marconi2004}. As discussed by \citet{2023Galax..11...21H}, only a portion of this bolometric luminosity is in the form of kinetic energy from AGN-driven winds. This fraction can be estimated by using the typical kinetic energy outflow rates from the multi-phase gas medium in samples of galaxies, with different studies finding values that range from $0.0001\% - 1\%\,L_{\rm{bol}}$ \citep[e.g.][see \citealt{2023Galax..11...21H} for a more extensive discussion]{Fiore2017,Kakkad2022,Dall'AgnoldeOliveira2021}. Given the large range of values observed from different studies for the different multi-phase mediums, here, for illustrative purposes, we take values in the range $0.1 - 0.5\%\,L_{\rm{bol}}$ and multiply this by the bolometric radiative energy density determined from \citet{Hopkins2007}, giving the range indicated by the green shaded region. We note that the upper end of this range is consistent with the median value of $0.5\%\,L_{\rm{bol}}$ used by \citet{2023Galax..11...21H}. These results indicate that AGN winds may have a comparable or even more important role in kinetic feedback at early cosmic times, but that the AGN jets may have been more important for feedback since $z \sim 2$.

\section{Conclusions}\label{sec:conclusions}
In this paper, we have presented the cosmic evolution of radio-AGN feedback out to $z \sim 2.5$, and compared these to predictions from recent cosmological-scale simulations. The observational data come from a sample of radio-AGN from the LoTSS Deep Fields, which consists of 9485 radio-excess AGN, of which 8409 are LERGs. The radio-excess AGN were identified based on sources with a $> 3\sigma$ excess in their radio emission over that expected from star-formation, and the LERGs identified as the subset of radio-excess AGN that do not show signs of having an accretion disk or torus structure (based on SED fitting of photometry). This dataset has been used previously to study the evolution of the radio luminosity functions of radio-excess AGN and LERGs by \citet{Kondapally2022}.

We studied the implications of the observed luminosity functions of radio-excess AGN, LERGs, and LERGs hosted by quiescent galaxies (Q-LERGs) on cosmic radio-AGN feedback. To do this, we translated the observed radio luminosities into jet kinetic powers using a scaling relationship. This was combined with the best-fitting models for the evolution of luminosity functions to compute the cosmic evolution of the kinetic luminosity density of radio-AGN (a proxy for the amount of heating provided by the radio-AGN jets) over cosmic time. We find that feedback from LERGs dominates the radio-mode feedback with the heating output for both the radio-excess AGN and LERGs being roughly constant across $0.5 < z \leq 2.5$, with both populations at $\Omega_{\rm{kin}} \approx 4-5 \times 10^{32}\,\rm{W\,Mpc^{-3}}$. At $z \lesssim 1$, most of the kinetic luminosity density of the total LERG population is output by Q-LERGs, beyond this redshift, the contribution from Q-LERGs decreases by an order of magnitude out to $z \sim 2.5$; this indicates that at higher redshifts star-forming galaxies contribute significantly to radio-AGN feedback. Our observations provide new measurements for the LERGs and Q-LERGs at $z \gtrsim 1$, while the measurements for the radio-AGN show good agreement with other studies in the literature. We also find that the relatively high-luminosity AGN dominate the overall heating output.

We then compared the observed evolution of the kinetic luminosity densities with predictions from various recent simulations. A comparison with the \textsc{sage} model finds that the Q-LERGs alone may deposit sufficient energy into their surrounding environment to balance the radiative cooling losses, providing evidence for a self-regulating AGN feedback cycle in these systems out to $z \sim 2.5$. In contrast, the mechanisms driving feedback in LERGs hosted by star-forming galaxies, which dominate at early times, remain unclear at present. The predictions from \textsc{radio-sage} for the LERGs show a good match to the observations; while the evolution of $\Omega_{\rm{kin}}(z)$ for the Q-LERGs from \textsc{radio-sage} shows a similar slope to the observations, it is offset to systematically lower values, suggesting a much higher contribution of heating output from LERGs in \textsc{radio-sage} coming from those hosted in star-forming galaxies. \textsc{simba} predicts $\Omega_{\rm{kin}}(z)$ for LERGs and Q-LERGs that agree with local observations at $z=0$, but rise out to $z \sim 1$ before declining at higher redshifts. The LERGs in \textsc{simba} over-predict the heating at $z \lesssim 1.5$, and under-predict it at higher redshifts compared to the observations. The Q-LERGs in \textsc{simba} also over-predict the heating output at $z \lesssim 1.5$ but match well with observations at higher redshifts. Although these differences are at the $0.3 - 0.4\,\rm{dex}$ level, these results suggest that in order to reproduce the observed massive galaxy population, the AGN jet feedback model employed in \textsc{simba} appears to require more heating output from the AGN across cosmic history than that observed.

We integrated the kinetic luminosity density across cosmic time to obtain the kinetic energy per unit volume output from AGN jets over cosmic history, $E_{\rm{kin}}$. Radio-excess AGN (and LERGs) output a total of $E_{\rm{kin}} \approx 10^{50}\,\rm{J\,Mpc^{-3}}$. The Q-LERGs are found to be the dominant source of this kinetic energy output in the form of AGN jets, with the Q-LERGs outputting most of their kinetic energy since $z \lesssim 0.5$. We compared this kinetic energy density from AGN jets to that output by AGN winds, estimated using measurements of the bolometric quasar luminosity function, to find that AGN jets dominate the total energy budget over AGN winds at $z \lesssim 2$; this indicates that AGN jets may play a more important role in kinetic feedback across much of cosmic history.

Our results indicate that while radio-AGN feedback plays an important role in galaxy evolution since at least $z \sim 2.5$, and simulations are broadly able to reflect this, black hole growth and AGN feedback models within simulations require modifications to reproduce the observed evolution of radio-AGN feedback in detail. To gain more insights into this, it is vital to not only study the total amount of feedback within simulations, but to also compare the feedback from different subsets of the radio-galaxy population, and the radio luminosity functions and host galaxy properties of different modes of radio-AGN; we will focus on studying this in a subsequent paper. In future, improvements to the observations will also be offered by deep high-resolution LOFAR imaging, enabling robust measures of source sizes out to high redshifts; this will enable improved dynamical modelling of the radio sources and a better-constrained jet power scaling relationship as a function of redshift.

\section*{Acknowledgements}
This paper is based (in part) on data obtained with the International LOFAR Telescope (ILT) under project codes LC0\_015, LC2\_024, LC2\_038, LC3\_008, LC4\_008, LC4\_034 and LT10\_01. LOFAR \citep{2013A&A...556A...2V} is the Low Frequency Array designed and constructed by ASTRON. It has observing, data processing, and data storage facilities in several countries, which are owned by various parties (each with their own funding sources), and which are collectively operated by the ILT foundation under a joint scientific policy. The ILT resources have benefitted from the following recent major funding sources: CNRS-INSU, Observatoire de Paris and Université d'Orléans, France; BMBF, MIWF-NRW, MPG, Germany; Science Foundation Ireland (SFI), Department of Business, Enterprise and Innovation (DBEI), Ireland; NWO, The Netherlands; The Science and Technology Facilities Council, UK; Ministry of Science and Higher Education, Poland. For the purpose of open access, the author has applied a Creative Commons Attribution (CC BY) licence to any Author Accepted Manuscript version arising from this submission. We thank the anonymous referee for their comments which have helped improve and clarify the paper.

RK and PNB acknowledge support from the UK Science and Technology Facilities Council (STFC) via grant ST/V000594/1. MJH and DJBS acknowledge support from the UK STFC [ST/V000624/1]. MB and IP acknowledge support from INAF under the SKA/CTA PRIN ``FORECaST'' and PRIN MAIN STREAM ``SAuROS'' projects, as well as through the Large Grant 2022 funding scheme (project ``MeerKAT and LOFAR Team up: a Unique Radio Window on Galaxy/AGN co-Evolution''). K.M. has been supported by the National Science Centre (UMO-2018/30/E/ST9/00082). LKM and NLT acknowledge support from the Medical Research Council [MR/T042842/1]. This research made use of {\sc Astropy}, a community-developed core Python package for astronomy \citep{astropy:2013, astropy:2018} hosted at \url{http://www.astropy.org/}, and of {\sc Matplotlib} \citep{hunter2007matplotlib}.

\section*{Data Availability}
The luminosity function observational data used in this study come from the analysis by \citet{Kondapally2022}, which can be accessed in a machine-readable format at the following repository: \url{https://github.com/rohitk-10/AGN_LF_Kondapally22}. The simulations data are based on analysis performed with the \textsc{simba} simulations \citep{dave2019simba,Thomas2021}, and with \textsc{radio-sage} \citep{Raouf2017}. The \textsc{simba} snapshots can be found at \url{http://simba.roe.ac.uk}. Other data presented in the paper are available upon request to the corresponding author.



\bibliographystyle{mnras}
\bibliography{lofar_agn_sims} 




\appendix
\section{Radio-AGN luminosity functions used for estimating kinetic heating rates}\label{sec:lf_comp}
\begin{table*}
\centering
\caption{\label{tab:reagn_fit}Results from modelling the evolution of three populations of AGN LFs, plotted in Fig.~\ref{fig:lerg_lf_comp}. The radio-excess AGN and LERG LFs are modelled by fitting a luminosity and density evolution (LDE) model at each redshift (independently) based on the local radio-AGN LF of \citet{Mauch2007}. For the Q-LERGs, the $\rho^{\star}(z)$ at each redshift is fixed based on the available quiescent host galaxies at that redshift, and the $L_{\star}(z)$ is then allowed to evolve freely at each redshift based on a broken power law fit to the observed LF at $0.5 < z \leq 1$ (with slopes of $\beta = -2.88$ and $\gamma = -0.55$; see \citealt{Kondapally2022}.)}
\begin{tabular}{cccc|ccc|ccc}
\hline\hline
$z$ & \multicolumn{3}{c|}{Radio-excess AGN} & \multicolumn{3}{c|}{LERGs} & \multicolumn{3}{c}{Quiescent LERGs}\Tstrut\Bstrut\\
{} & $\log_{10} \rho^{\star}(z)$ & $\log_{10} L_{\star}(z)$ & $\chi^{2}_{\nu}$ & $\log_{10} \rho^{\star}(z)$ & $\log_{10} L_{\star}(z)$ & $\chi^{2}_{\nu}$ & $\log_{10} \rho^{\star}(z)$ & $\log_{10} L_{\star}(z)$ & $\chi^{2}_{\nu}$\Bstrut\\
\hline
${0.5} < z \leq {1.0}$ & $-5.53_{-0.08}^{+0.07}$ & $26.31_{-0.12}^{+0.14}$ & 7.22 & $-5.39_{-0.07}^{+0.07}$ & $25.98_{-0.11}^{+0.12}$ & 5.59 & $-6.37_{-0.24}^{+0.17}$ & $27.03_{-0.22}^{+0.40}$ & 3.57\Tstrut\\[0.15cm]
${1.0} < z \leq {1.5}$ & $-5.89_{-0.09}^{+0.08}$ & $26.77_{-0.14}^{+0.16}$ & 5.83 & $-5.63_{-0.09}^{+0.08}$ & $26.24_{-0.13}^{+0.14}$ & 5.56 & -6.92 & $27.33_{-0.04}^{+0.04}$ & 3.39\\[0.15cm]
${1.5} < z \leq {2.0}$ & $-5.71_{-0.08}^{+0.07}$ & $26.76_{-0.12}^{+0.13}$ & 7.98 & $-5.52_{-0.08}^{+0.08}$ & $26.35_{-0.11}^{+0.12}$ & 6.28 & -7.28 & $27.91_{-0.06}^{+0.06}$ & 2.28\\[0.15cm]
${2.0} < z \leq {2.5}$ & $-5.41_{-0.09}^{+0.09}$ & $26.09_{-0.13}^{+0.13}$ & 4.98 & $-5.13_{-0.11}^{+0.12}$ & $25.61_{-0.14}^{+0.14}$ & 3.45 & -7.61 & $27.93_{-0.12}^{+0.12}$ & 5.89\\[0.15cm]
\hline
\end{tabular}
\end{table*}

\begin{figure*}
    \centering
    \includegraphics[width=\textwidth]{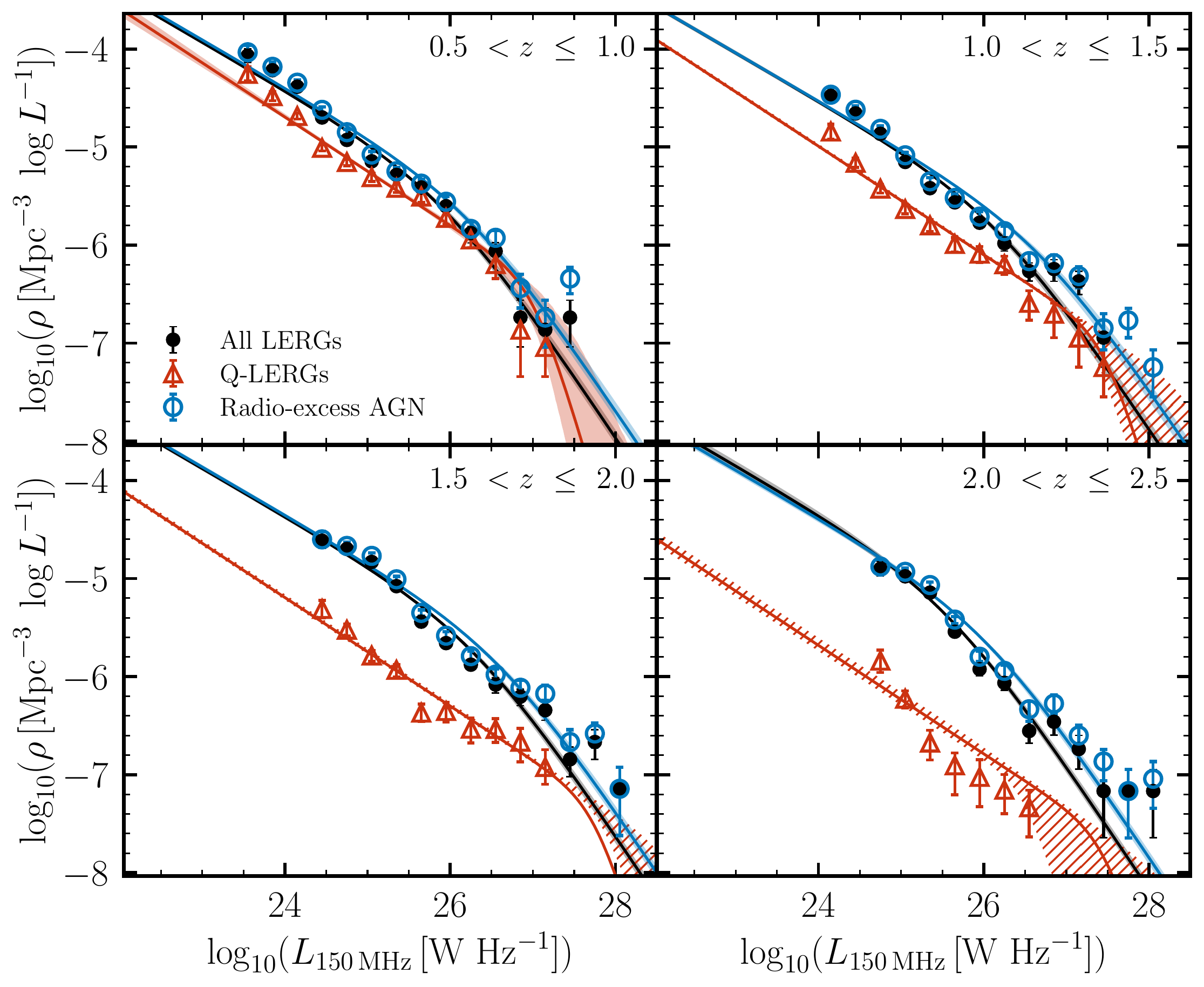}
    \caption{\label{fig:lerg_lf_comp}Cosmic evolution of AGN luminosity functions from LoTSS-Deep across $0.5 < z \leq 2.5$. The LFs for radio-excess AGN (blue open circles), LERGs (black circles), and quiescent LERGs (Q-LERGs; red triangles) are shown. The corresponding solid lines at each redshift show the best-fitting LFs for the three populations, in which the lower-$z$ LF shape (see text) is evolved in both density and luminosity (the shaded regions represent the uncertainty on this fit; see text).}
\end{figure*}

The observed radio LFs used in this study, for the radio-excess AGN, LERGs, and Q-LERGs are taken from the analysis by \citet{Kondapally2022}. In summary, the LFs were calculated using the $1/V_{\rm{max}}$ method, which accounted for the radio flux-density incompleteness and noise variation across the field. The radio flux-density completeness corrections were estimated on a field-by-field basis by performing a large suite of simulations involving the injection and recovery of mock sources onto the image plane; we refer to \citet{Kondapally2022} for the details of this process. The 150\,MHz radio luminosities were computed assuming a spectral index $\alpha = -0.7$ \citep[e.g.][]{CalistroRivera2017,Murphy2017}.

\citet{Kondapally2022} computed the LERG and Q-LERG LFs over the redshift range: $0.5 < z \leq 1$, $1 < z \leq 1.5$, $1.5 < z \leq 2$, and $2 < z \leq 2.5$. They computed the radio-excess AGN LFs over different redshift bins to readily compare with available observational measurements for that population. Here, we recompute the radio-excess AGN LFs over the same redshift bins as that for the LERGs, which are shown in Fig.~\ref{fig:lerg_lf_comp}. Both the radio-excess AGN LFs and LERG LFs show excellent agreement with the observed LFs from other studies \citep{2014MNRAS.445..955B,smolcic2017agn_evol_vla,Butler2019} over the full redshift range studied as shown by \citet{Kondapally2022}. From Fig.~\ref{fig:lerg_lf_comp}, it is also evident that across all radio luminosities and redshifts studied, the LoTSS-Deep radio-excess AGN sample is dominated by LERGs; the HERGs form a minority of the radio-excess AGN population. As noted by \citet{Kondapally2022}, we observe little-to-no evolution in the LERG LFs across $0.5 < z \leq 2.5$, however when split by host galaxy type, we see a strong decline in the space densities of the quiescent-LERGs (Q-LERGs, red triangles) with increasing redshift. \citet{Kondapally2022} showed that this decline is in line with the decrease in the availability of quiescent hosts.

The evolution of the radio-AGN LFs were modelled at each redshift as a broken power law of the form
\begin{equation}\label{eq:bpl}
\rho(L,z) = \frac{\rho^{\star}(z)}{\left( L^{\star}(z)/L\right)^{\beta} + \left( L^{\star}(z)/L\right)^{\gamma}},
\end{equation}
where $L^{\star}(z)$ is the characteristic luminosity at redshift $z$, $\rho^{\star}$ is the characteristic space density at redshift $z$, and $\beta$ and $\gamma$ are the bright and faint end slopes, respectively. For both the radio-excess AGN and the LERGs, a broken power-law fit at each redshift was performed by fixing the bright and faint end slopes, $\beta=-1.27$ and $\gamma=-0.49$, respectively, following the broken power-law fit to the local radio-AGN LF by \citet{Mauch2007}; we then fitted for a combined luminosity and density evolution (LDE) model at each redshift, as detailed by \citet{Kondapally2022}. The resulting fitted $\rho^{\star}(z)$ and $L^{\star}(z)$ for LERGs (taken from \citealt{Kondapally2022}) and radio-excess AGN are shown in Table~\ref{tab:reagn_fit}. For the Q-LERGs, we directly fitted a broken power law to the $0.5 < z \leq 1.0$ LF to derive empirical power law slopes. Using these slopes, we fitted the evolution at higher redshifts by fixing $\rho^{\star}(z)$ at each redshift bin in accordance with the evolution of the available quiescent host-galaxies and only fitting for $L^{\star}(z)$ (see \citealt{Kondapally2022} for details); the parameters for the best-fitting Q-LERG LFs are also shown in Table~\ref{tab:reagn_fit} for completeness.

\section{Comparison of jet power inference methods}\label{ap:lkin_rels}
\begin{table*}
    \caption{\label{tab_ap:lkin_rels}Different literature scaling relations for converting monochromatic radio luminosities ($L_{1.4\,\rm{GHz}}$) to jet kinetic powers ($L_{\rm{kin}}$) that are compared. For the \citet{Willott1999} and \citet{2014ARA&A..52..589H}, the relations listed have explicitly assumed $f_{\rm{W}} = 15$ and $f_{\rm{cav}} = 4$, as used in this study; relations for other factors can be calculated using the additional term in each case. The \citet{Shabala2013} relation includes a dependence on redshift and source size.}
    \centering
    \begin{tabular}{ccc}
    \hline\hline
    Reference & Scaling Relation \\
    \hline
    \citet{Willott1999} & $\log_{10}(L_{\rm{kin}}/\rm{W}) = 0.86 \log_{10}\left( \frac{L_{\rm{1.4\,GHz}}}{10^{25}\, \rm{W\,Hz^{-1}}} \right) + 37.37 + \log_{10} \left(\frac{\mathit{f}_{\rm{W}}}{15}\right)$ \Tstrut\Bstrut \\[0.15cm]
    \citet{Kording2008} & $\log_{10}(L_{\rm{kin}}/\rm{W}) = 0.71 \log_{10}\left(\frac{L_{\rm{1.4\,GHz}}}{10^{25}\, \rm{W\,Hz^{-1}}}\right) + 37.72$\\[0.15cm]
    \citet{Cavagnolo2010} & $\log_{10}(L_{\rm{kin}}/\rm{W}) = 0.75 \log_{10}\left(\frac{L_{\rm{1.4\,GHz}}}{10^{25}\, \rm{W\,Hz^{-1}}}\right) + 38.06$\\[0.15cm]
    \citet{O'Sullivan2011} & $\log_{10}(L_{\rm{kin}}/\rm{W}) = 0.63 \log_{10}\left(\frac{L_{\rm{1.4\,GHz}}}{10^{25}\, \rm{W\,Hz^{-1}}}\right) + 37.76$\\[0.15cm]
    {\citet{Shabala2013}} & $\log_{10}(L_{\rm{kin}}/\rm{W}) = 0.8\log_{10}\left( \frac{L_{\rm{1.4\,GHz}}}{10^{25}\, \rm{W\,Hz^{-1}}} \right) + 35.12 + \log_{10}(1+\textit{z}) + 0.58\log_{10}\left(\frac{\textit{D}}{\rm{kpc}}\right)$ \\[0.15cm]
    {\citet{2014ARA&A..52..589H}} & $\log_{10}(L_{\rm{kin}}/\rm{W}) = 0.68 \log_{10}\left( \frac{L_{\rm{1.4\,GHz}}}{10^{25}\, \rm{W\,Hz^{-1}}} \right) + 37.45 + \log_{10} \left(\frac{\mathit{f}_{\rm{cav}}}{4}\right)$ \\[0.15cm]
    \citet{Ineson2017} & $\log_{10}(L_{\rm{kin}}/\rm{W}) = 0.89 \log_{10}\left( \frac{L_{\rm{1.4\,GHz}}}{10^{25}\, \rm{W\,Hz^{-1}}} \right) + 36.05$\\[0.15cm]
    \hline
    \end{tabular}
\end{table*}
\begin{figure}
    \centering
    \includegraphics[width=\columnwidth]{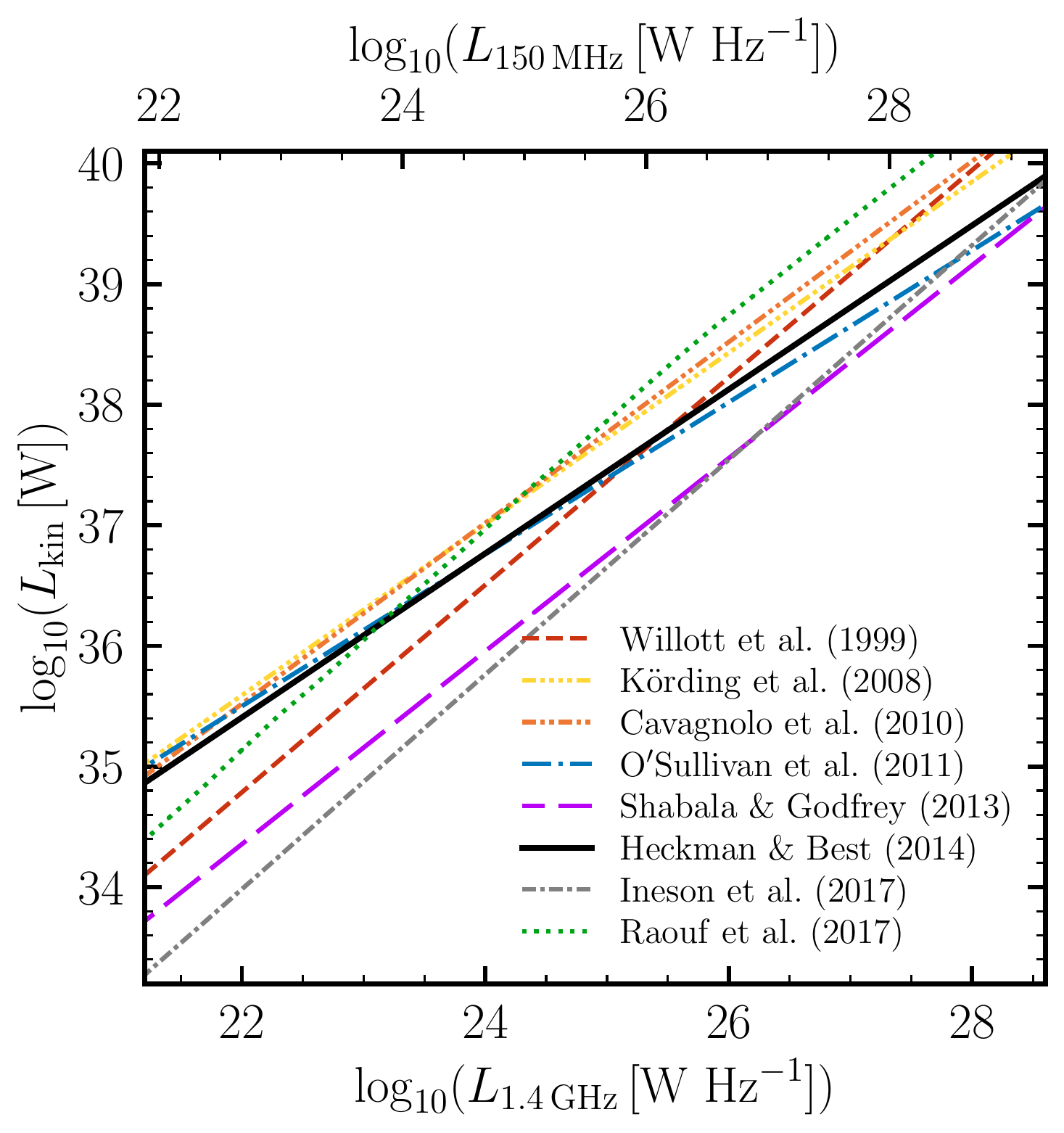}
    \caption{\label{fig_ap:lkin_lum_rels}Comparison of different scaling relations between monochromatic radio luminosity (1.4\,GHz) and the jet powers in the literature, as highlighted in Table~\ref{tab_ap:lkin_rels} \citep[][]{Willott1999,Kording2006,Kording2008,Cavagnolo2010,O'Sullivan2011,2014ARA&A..52..589H,Godfrey2016,Ineson2017,Raouf2017}. The 150\,MHz radio luminosities (assuming $\alpha = -0.7$) are also shown. The \citet{2014ARA&A..52..589H} relation is shown for $f_{\rm{cav}} = 4$, which shows good agreement in its normalisation with that of \citet{Willott1999} when using $f_{\rm{W}} = 15$, and with the \citet{O'Sullivan2011} relation. For illustrative purposes, the \citet{Shabala2013} relation is shown for a source of size $D = 100\,\rm{kpc}$ and $z=2$. The median relation for the jet model from \textsc{radio-sage} is taken from \citet{Raouf2017} and extrapolated linearly beyond $\sim 10^{26}\,\rm{W\,Hz^{-1}}$, which shows a higher normalisation than the other relations considered.}
\end{figure}

There exist many scaling relations for translating monochromatic radio luminosities ($L_{\rm{1.4\,GHz}}$) into jet kinetic powers ($L_{\rm{kin}}$) derived using different methods \citep[e.g.][]{Willott1999,Birzan2004,Cavagnolo2010,O'Sullivan2011,Shabala2013,Godfrey2016}, each with their own set of uncertainties on the calibration \citep[see][]{2014MNRAS.445..955B,Godfrey2016,Hardcastle2020}. In this section, we compare the commonly used scaling relations and assess their impacts on the cosmic kinetic luminosity density results presented in Section~\ref{sec:heating_rate}.

Fig.~\ref{fig_ap:lkin_lum_rels} shows the different scaling relations between 1.4\,GHz radio luminosity and jet kinetic powers. These scaling relations are also listed in Table~\ref{tab_ap:lkin_rels}. Throughout the main sections of this paper, we have used the scaling relation from \citet{2014ARA&A..52..589H}; this method is based on determining the work done by the radio jets in inflating cavities using X-ray observations \citep[e.g.][]{Birzan2004,Birzan2008,Cavagnolo2010}. \citet{O'Sullivan2011} and \citet{2014ARA&A..52..589H} extended the analysis to fainter radio luminosities, albeit for a small number of systems in total at low redshift ($z < 0.04$), and found slopes $\sim 0.6 - 0.8$ with a typical scatter of $\sigma \approx 0.7\,\rm{dex}$.

Another commonly used relation in the literature is that based on theoretical arguments; \citet{Willott1999} used the minimum energy argument to estimate the energy stored within radio lobes using estimates of radio source lifetimes and the efficiency with which the jet power is converted to synchrotron emission as the radio lobes expand, as described in Section~\ref{sec:heating_rate}. They combined all the uncertainties (departure from minimum energy, jet plasma composition, and energy stored in radiating particles) in their calibration into a single parameter, $f_{\rm{W}}$ which is expected to lie in the range $f_{\rm{W}} = 1 - 20$. \citet{Willott1999} find a steeper slope ($\sim$ 0.9) than the cavity power based methods, but for the typical value of $f_{\rm{W}} = 15$ \citep[e.g.][]{2014ARA&A..52..589H,smolcic2017agn_evol_vla}, this relation produces a similar normalisation to that of the cavity based estimates at typical luminosities of $L_{\rm{1.4\,GHz}} \sim 10^{25}\,\rm{W\,Hz^{-1}}$ \citep{2014ARA&A..52..589H}. More recently, \citet{Ineson2017} estimated the jet powers using internal energies of the lobes for a sample of bright Fanaroff-Riley II (FR-II) sources ($L_{\rm{1.4\,\rm{GHz}}} \gtrsim 10^{24}\,\rm{W\,Hz^{-1}}$), finding a similar slope to the \citet{Willott1999} relation with a normalisation that is consistent with a lower $f_{\rm{W}}$ value of $f_{\rm{W}} \sim 4$ (see Table~\ref{tab_ap:lkin_rels}).

\citet{Shabala2013} used a sample of FR-II sources to derive a jet-power scaling relationship based on both the radio luminosity and the source-size (which is taken to be a proxy for the age of the radio source). In Fig.~\ref{fig_ap:lkin_lum_rels}, we show the \citet{Shabala2013} relation assuming a source-size of 100\,kpc at $z=2$; this relation predicts lower jet powers than the other relations, but is found to be consistent with the \citet{Willott1999} relation when assuming $f_{\rm{W}} \sim 4$.

\begin{figure*}
    \centering
    \includegraphics[width=0.8\textwidth]{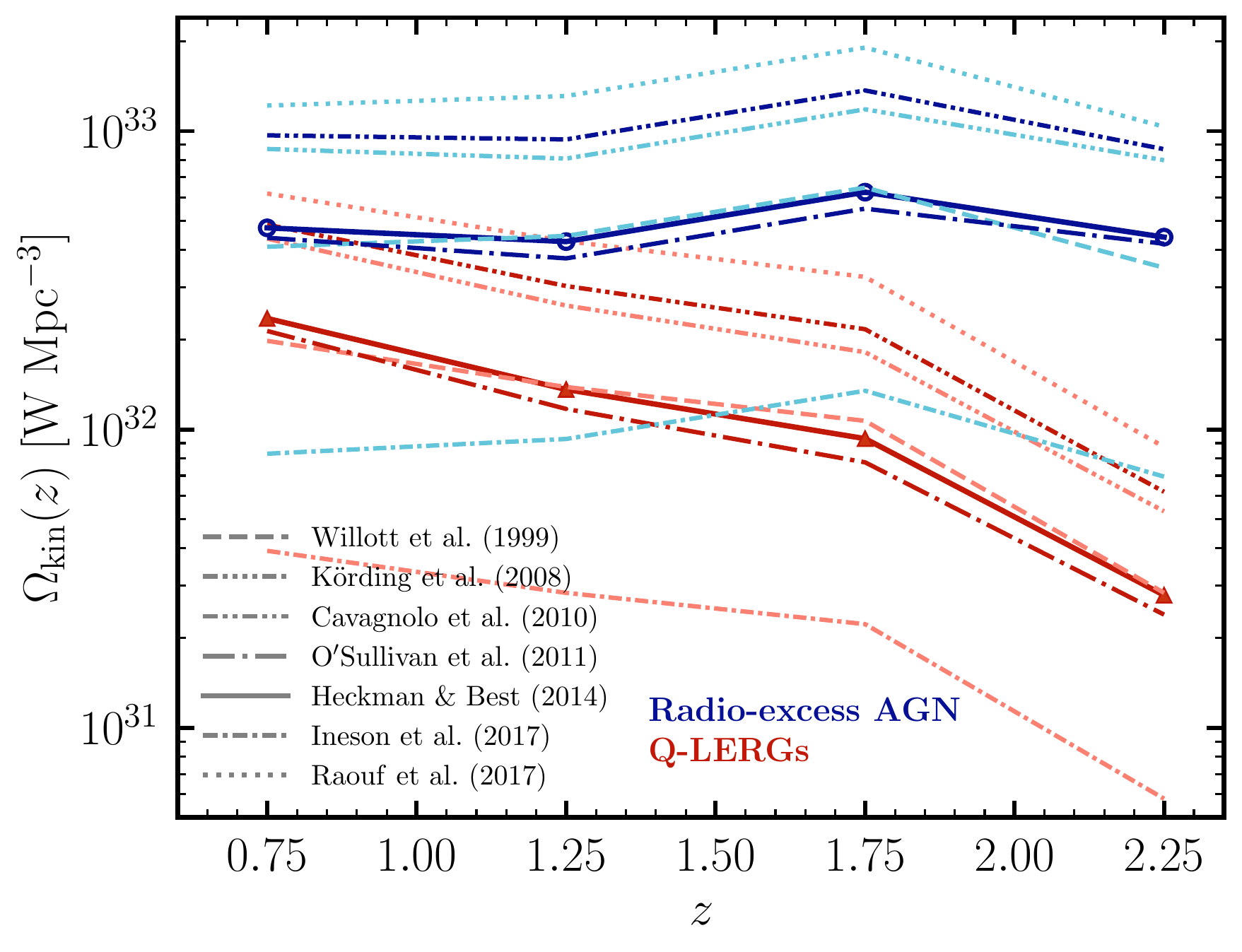}
    \caption{\label{fig_ap:hrate_z_var}The cosmic evolution of the kinetic luminosity density for radio-excess AGN (blue) and Q-LERGs (red) assuming different scaling relations for converting monochromatic radio luminosities into jet powers in Fig.~\ref{fig_ap:lkin_lum_rels}. The darker shades of colours correspond to measurements based on cavity power method and the lighted shades correspond to methods based on the synchrotron emission.}
\end{figure*}

Also shown in Fig.~\ref{fig_ap:lkin_lum_rels} is the median relationship found for the jet model employed within \textsc{radio-sage} \citep{Raouf2017}; this has a notably steeper slope than the \citet{2014ARA&A..52..589H} and \citet{O'Sullivan2011} relation, similar to that of the \citet{Willott1999} relation, however with a higher normalisation. Therefore, \textsc{radio-sage} AGN on average would be expected to produce higher kinetic powers at high luminosities and hence result in higher kinetic luminosity density across redshift (see Fig.~\ref{fig_ap:hrate_z_var}). The \citet{Kording2008} relation for jet powers based on observations of X-ray binaries is also shown; this relation has a similar slope to that of \citet{2014ARA&A..52..589H}, however with a higher normalisation (by $\sim 0.3\,\rm{dex}$) such that it predicts higher kinetic powers than the \citet{2014ARA&A..52..589H} scaling relation used in this study over all radio luminosities.

In Fig.~\ref{fig_ap:hrate_z_var}, we show the cosmic evolution of the kinetic luminosity density of AGN within LoTSS-Deep (see Sect.~\ref{sec:hrate_evol}) when assuming different scaling relations for converting radio luminosities to kinetic jet powers. The solid lines with symbols (the same as in Fig.~\ref{fig:lerg_hrate}) show the measurements based on the \citet{2014ARA&A..52..589H} relation that is used in this study (see Section~\ref{sec:heating_rate}) for radio-excess AGN (blue) and Q-LERGs (red); the `All-LERGs' are not shown on this plot for clarity. Other line styles show how the kinetic luminosity density changes for these two populations when different scaling relations are used for the two AGN populations.

Measurements based on cavity power scaling relations \citep[i.e. from][]{Cavagnolo2010,O'Sullivan2011,2014ARA&A..52..589H} are shown by darker colour shades compared to estimates based on radio source modelling by \citet{Willott1999,Kording2008,Ineson2017,Raouf2017}. The cavity power based estimates from \citet{O'Sullivan2011} and \citet{2014ARA&A..52..589H} agree well with each other and with the results from \citet[][for the value of $f_{\rm{W}}=15$]{Willott1999}. The results from \citet{Cavagnolo2010} predict systematically higher kinetic luminosity densities than the other cavity based methods (by a factor of $\sim$ 2) due to the steeper slope; the \citet{Cavagnolo2010} sample of 21 systems covered relatively low radio luminosities ($L_{\rm{1.4\,GHz}} \lesssim 10^{24}\,\rm{W\,Hz^{-1}}$) compared to both \citet{O'Sullivan2011} and \citet{2014ARA&A..52..589H} resulting in poorer constraints and hence extrapolation at high luminosities. This can have an important effect as the heating output peaks near the break in the LFs, i.e. at $L_{\rm{150\,MHz}} \sim 10^{26} - 10^{27}\,\rm{W\,Hz^{-1}}$ (see Fig.~\ref{fig:hrate_l150}). Both the \citet{Kording2008} and \citeauthor{Raouf2017} (\citeyear{Raouf2017}; \textsc{radio-sage}) models predict systematically higher $\Omega_{\rm{kin}}(z)$ than the \citet{2014ARA&A..52..589H} relation used in this study. The shape of the cosmic evolution seen for the various relations used is similar, with the different relations largely introducing a difference in overall normalisation of the kinetic luminosity densities.


\bsp	
\label{lastpage}
\end{document}